\documentclass[12pt]{article}

\usepackage{graphics}
\usepackage{graphicx}
\usepackage{feynmf}
\usepackage{color}
\usepackage{subfigure}
\usepackage[utf8]{inputenc}
\usepackage[english]{babel}
\usepackage{amsmath}
\usepackage{amsfonts}
\usepackage{amssymb}
\usepackage{epsfig}
\usepackage{graphics,psfrag,rotating}
\usepackage{graphicx}
\usepackage{dcolumn}
\usepackage{bm}
\usepackage{epstopdf}
\newcommand{\be}{\begin{equation}}
\newcommand{\ee}{\end{equation}}
\newcommand{\ba}{\begin{eqnarray}}
\newcommand{\ea}{\end{eqnarray}}

\newcommand{\bw}{\begin{widetext}}
\newcommand{\ew}{\end{widetext}}

\usepackage{fancyhdr}

\newlength\encuadernacion \newlength\longB 
\newenvironment{narrow}[2]{%
\begin{list}{}{%
\setlength{\topsep}{0pt}%
\setlength{\leftmargin}{#1}%
\setlength{\rightmargin}{#2}%
\setlength{\listparindent}{\parindent}%
\setlength{\itemindent}{\parindent}%
\setlength{\parsep}{\parskip}}%
\item[]}{\end{list}}

\begin{document}
\input epsf

\fancypagestyle{plain}{
\fancyhf{}
\fancyfoot[C]{\thepage}
\renewcommand{\headrulewidth}{0pt}
\renewcommand{\footrulewidth}{0pt}}

\pagestyle{plain}

\begin{center}
\large{This essay received an{\it ``Honorable Mention''}
in the\\Gravity Research Foundation 2014 Awards for Essays on Gravitation\\}
\vspace{1cm}
\huge{\bf The large-scale structure of vacuum} \\
\vspace*{1cm}
\large{\bf F.\,D.\;Albareti$^{a}$, J.\,A.\,R.\;Cembranos$^{b}$ and A.\,L.\;Maroto$^{b}$}  \\
\vspace{0.75cm}
\normalsize
$^a$Departamento de F\'isica Te\'orica,
Universidad Aut\'onoma de Madrid,\\
Campus de Cantoblanco, 28049 Madrid, Spain. \\
\vspace{0.5cm}
$^b$Departamento de F\'{\i}sica Te\'orica,
Universidad Complutense de Madrid,\\
Ciudad Universitaria, 28040 Madrid, Spain.
\vspace{0.5cm}

March $31^{\text{st}}$, 2014

\vspace*{0.75cm} {\bf ABSTRACT}\\  \end{center} \begin{narrow}{1cm}{1cm}

The vacuum state in quantum field theory is known to exhibit an important number
of  fundamental physical features. In this work we explore the possibility that
this state could also present a non-trivial space-time structure on large scales.
In particular, we will show that by imposing the renormalized vacuum energy-momentum
tensor to be conserved and compatible with cosmological observations,
the vacuum energy of sufficiently heavy fields
behaves at late times as non-relativistic matter rather than as a cosmological constant. In this limit, the vacuum state supports perturbations
whose speed of sound is negligible and accordingly allows the growth of structures in the vacuum energy itself. This large-scale
structure of vacuum could seed the formation of galaxies and clusters very much in the same
way as cold dark matter does.


\end{narrow}

\vspace*{0.5cm}
\vfill
\noindent
\rule[.1in]{4.5cm}{.002in}\\
\noindent E-mail: franco.albareti@uam.es,  cembra@ucm.es, maroto@ucm.es

\newpage
\baselineskip 20pt


\fancypagestyle{plain}{
\fancyhf{}
\fancyfoot[C]{\thepage}
\fancyhead[C]{{\it The large-scale structure of vacuum} -- Essay on Gravitation for the GRF 2014 Awards}
\renewcommand{\headrulewidth}{0.75pt}
\renewcommand{\footrulewidth}{0pt}}

\pagestyle{plain}

With the advent of quantum field theories, the classical concept of vacuum as the absence of any kind of matter or energy changed
dramatically. Heisenberg's uncertainty principle not only
demands the presence of a minimum energy per field mode, giving rise to a non-vanishing (and divergent)
vacuum energy, but also those vacuum fluctuations are known to modify fundamental properties
 of the own fields such as their charges or masses. Moreover, the quantum vacuum exhibits a
rich structure regarding electroweak and color interactions. Thus we know that the
QCD vacuum is characterized  by  non-vanishing
expectation values of the gluon $\langle \alpha_s G_{\mu\nu}^aG^{\mu\nu\,a}/\pi\rangle \simeq (300\,\text{MeV})^4$
and quark $\langle \bar \psi \psi\rangle \simeq -(230\,\text{MeV})^3$ condensates. The later has a fundamental
physical consequence since it is responsible for the chiral symmetry breaking.
Other features related to the QCD contribution to the vacuum state have not been detected yet. Indeed, the non-observation
of the effects corresponding to its non-trivial topological structure imposes very stringent
upper limits on the actual value of the $\theta$  parameter associated to the strong CP problem.

Electroweak vacuum  is also known to present a non-vanishing expectation value for the Higgs field
$\langle H\rangle\simeq 250 \,\text{GeV}$, which is responsible
for the electroweak symmetry breaking and the generation of the masses of the rest of Standard Model
particles.

These  expectation values characterizing the non-trivial properties of the
vacuum are usually calculated in Minkowski space-time and after a regularization and
renormalization process they are just given by constant parameters in space-time
whose values are fixed experimentally. However, in the presence of a 
background metric, those expectation values could acquire a
non-trivial space-time dependence \cite{Birrell}.
In this work we are precisely interested in obtaining such a dependence in the particular case of
vacuum energy in a cosmological context. Thus, we will compute the vacuum energy of a simple massive scalar field model in a
perturbed Robertson-Walker background. We will show that, demanding the renormalized vacuum energy-momentum tensor
to be covariantly conserved and the value of the renormalized vacuum energy to be  compatible
with observations, there is a contribution as non-relativistic matter that dominates at late
times. This contribution is shown to support perturbations whose speed of sound is
negligible and accordingly allows the growth of structures in the vacuum energy itself. This large-scale
structure of vacuum could seed the formation of galaxies and clusters very much in the same
way as cold dark matter does.

Let us consider  a scalar field $\phi$ with mass $m$ in a spatially flat Robertson-Walker background
$\text{d}s^2=a^2(\eta)(\text{d}\eta^2-\text{d}{\bf x}^2)$, the corresponding Klein-Gordon equation reads
\begin{eqnarray}
\Box \, \phi +m^2\, \phi = 0 \,.
\label{KG}
\end{eqnarray}
The field $\phi(\eta,{\bf x})$ can be Fourier expanded as
\begin{eqnarray}
\phi (\eta,{\bf x})=\int \text{d}^3{\bf k}\, \left(a_{{\bf k}}\, \phi_{k}(\eta)\,e^{i\, {\bf k}\,{\bf x}} + a^{\dag}_{{\bf k}}\, \phi^*_{k}(\eta)\,e^{-i\, {\bf k}\,{\bf x}}\right)\,,
\label{fielddecomposition}
\end{eqnarray}
where the  $a_{\bf k}$ and $a^{\dag}_{{\bf k}}$ operators satisfy the usual commutation relations
\begin{eqnarray}
[a_{{\bf p}},a^{\dag}_{{\bf q}}]=\delta^{(3)}({\bf p}-{\bf q})\,.
\end{eqnarray}
Introducing $\psi_{k}$ by
\begin{eqnarray}
\phi_{k} = \frac{\psi_{k}}{a}\,,
\label{mode}
\end{eqnarray}
eq.\ \eqref{KG} can be recast as
\begin{eqnarray}
\psi_{k}'' + \left(k^2 - \frac{a''}{a} + m^2 a^2 \right)\psi_{k}=0\,
\label{fieldequation}
\end{eqnarray}
where primes stand for derivative with respect to conformal time.

The non-vanishing  vacuum expectation values of the energy-momentum tensor components are \cite{Parker,Fulling}
\begin{eqnarray}
\rho=\langle 0\vert T^{0}_{\;0}\vert 0\rangle&=& \int \frac{\text{d}^3{\bf k}}{2a^2} \left(\vert{\phi'_{k}}\vert^2+k^2\vert\phi_{k}\vert^2
+m^2a^2\vert\phi_{k}|^2\right)\,
\end{eqnarray}
\begin{eqnarray}
p=-\langle 0\vert T^{i}_{\;i}\vert 0\rangle&=& \int \frac{\text{d}^3{\bf k}}{2a^2} \left(\vert{\phi'_{k}}\vert^2-\frac{ k^2}{3}\vert\phi_{k}\vert^2
-  m^2a^2\vert\phi_{k}|^2\right)\,.
\end{eqnarray}
In order to obtain the corresponding renormalized quantities, we will use
a constant comoving cutoff over the three-momenta \cite{VDM}, i.e
we will consider
\begin{eqnarray}
\rho_{\text{ren}}&=& \frac{2\pi}{a^2}\int_0^{\Lambda_R} \text{d}{k \,k^2} \left(\vert{\phi'_{k}}\vert^2+k^2\vert\phi_{k}\vert^2
+m^2a^2\vert\phi_{k}|^2\right)\,
\label{energydensity}
\end{eqnarray}
\begin{eqnarray}
p_{\text{ren}}&=&  \frac{2\pi}{a^2}\int_0^{\Lambda_R} \text{d}{k \,k^2} \left(\vert{\phi'_{k}}\vert^2-\frac{ k^2}{3}\vert\phi_{k}\vert^2
-  m^2a^2\vert\phi_{k}|^2\right)\,,
\label{pressure}
\end{eqnarray}
where as for any other renormalized quantity in field theory, the actual value of $\Lambda_R$ can only be determined by observations.
 Notice that unlike Minkowski space-time \cite{Akhmedov,Ossola}, the constant three-momentum cutoff does not
break the symmetries of the underlying Robertson-Walker metric. In fact, the use of a comoving cutoff guarantees the renormalized energy-momentum tensor to be covariantly conserved.

We shall be interested in the equation of state of vacuum energy $w=p_{\text{ren}}/\rho_{\text{ren}}$.
For that purpose it is important to notice that there are two (comoving)
mass scales in the theory, namely,  $ma$ and $\Lambda_R$. Thus, in the UV limit ($\Lambda_R\gg ma$),
which is the limit usually considered in the literature \cite{Martin,Maggiore}, it can be seen that there are three types of  contributions to
$\rho_{\text{ren}}$, namely,
terms scaling as $\Lambda_R^4/a^4$, $m^2\Lambda_R^2/a^2$ and $m^4\log \Lambda_R$,
where for simplicity we have assumed $m^2\gg H^2$, which is actually the case for the current value of the Hubble parameter $H$ and the masses of the SM fields. Each of them are independently conserved and could
be removed also independently by adding appropriate conserved counterterms to the
energy-momentum tensor. In particular,
by eliminating the quartic and quadratic terms we would obtain the standard cosmological
constant contribution that scales logarithmically with the cutoff.

However, the value of $\Lambda_R$ is severely constrained by entropic arguments in order to avoid
overcounting the number of states accessible to a gravitating system. These holographic bounds
suggest that the correct limit for the computation of the contribution to the vacuum energy may
be the opposite \cite{holo1,holo2,holo3}.
Indeed, in the IR limit ($\Lambda_R\ll ma$), we will show that the dominant contribution always scales as
$m\Lambda_R^3/a^3$ regardless the (conserved) counterterms which are added
to renormalize the UV behaviour. Notice that for sufficiently large masses and/or  late times
this will be the dominant contribution to the vacuum energy.

\begin{figure}

        {\begin{center}    \includegraphics[width=0.48\textwidth]{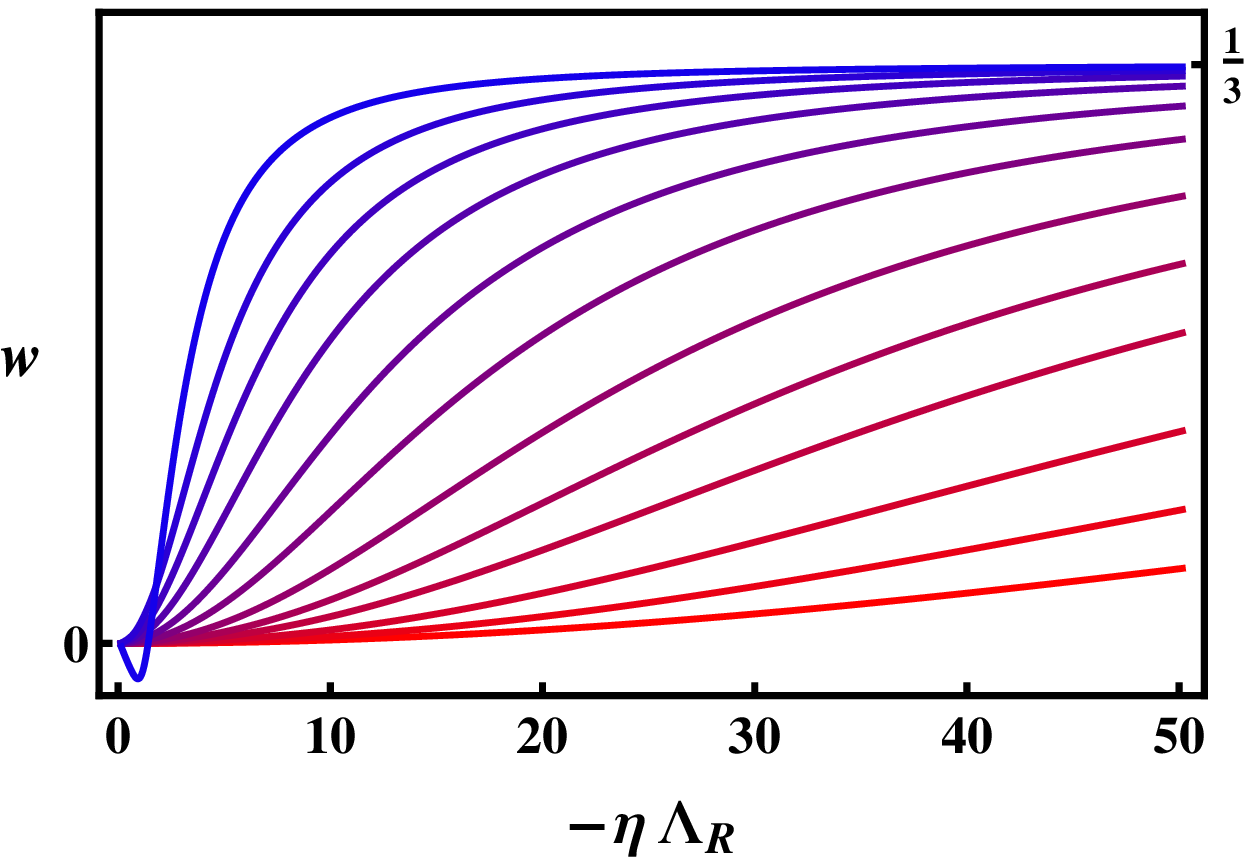}
        \includegraphics[width=0.5\textwidth]{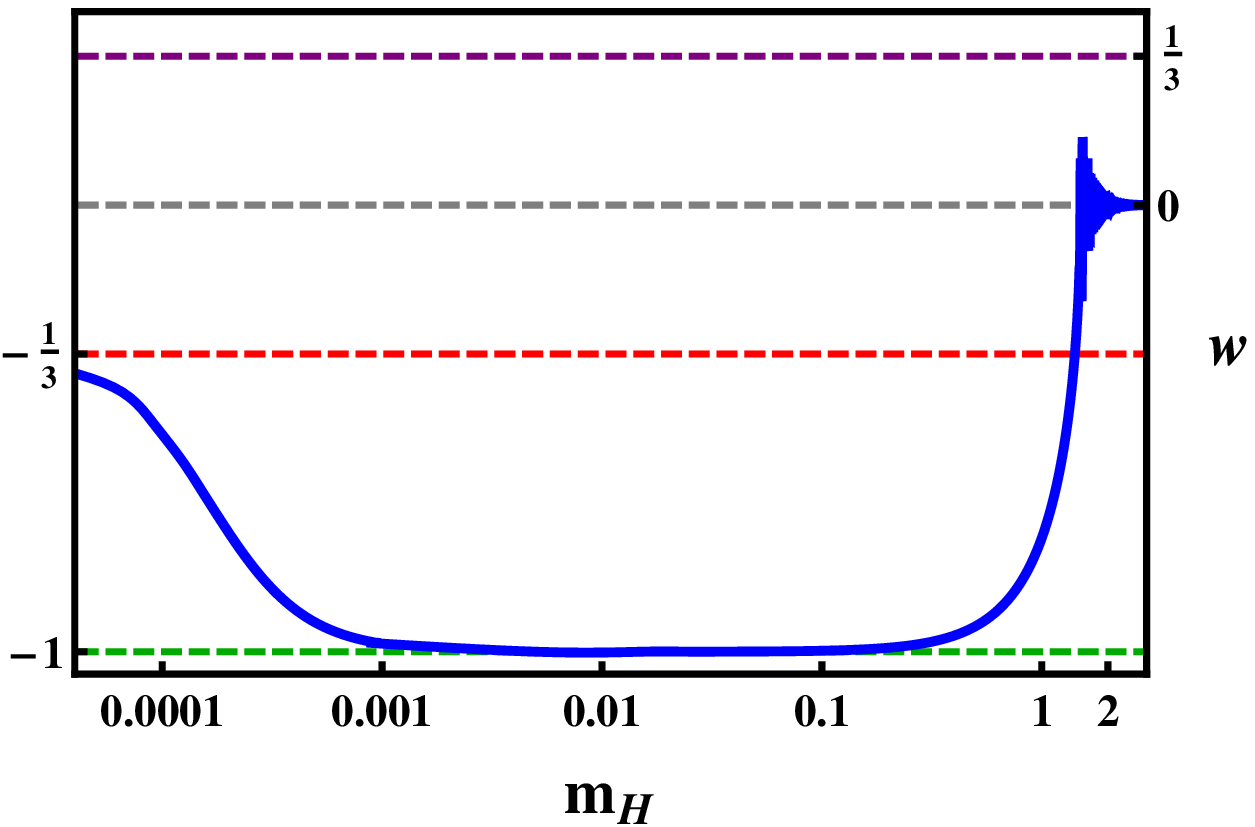}    \end{center}
    {\footnotesize Left: Exact evolution of the equation of state parameter $w$ in a de Sitter space-time \cite{VDM} in terms of the conformal time $\eta$ and the renormalization cutoff $\Lambda_{R}$ for several values of the field mass $m_{H}$ measured in units of $H$ (from top to bottom $m_{H}=2, 3.5, 5, 7, 10, 14, 20, 27, 35, 50, 70, 100$) (in linear-linear scale). For past times (or high cutoff) the dominant contribution is the radiation one $w=1/3$. As the time goes by (or for low cutoff), a matter behavior $w=0$ appears. Right: Linear-log scale plot of the equation of state parameter $w$ when $-\eta\,\Lambda_{R}=0.1$ for field masses in the range $3\cdot10^{-5}<m_{H}<2.2$. For $m\gg H$, $w=0$ and a non-relativitic matter behaviour is found.\\}
}
\end{figure}

Let us then consider the case $\Lambda_R\ll ma$ with $m\gg H$.  Equation \eqref{fieldequation} is reduced in this case to
\begin{eqnarray}
\psi_{k}'' + m^2 a^2 \psi_{k}=0\,,
\end{eqnarray}
whose approximated solution can be obtained by the WKB method \cite{Birrell}
\begin{eqnarray}
\phi _k(\eta)&=& \frac{c_{1}(k)}{\sqrt{2ma^3}} \exp{\left(-im\int a \,\text{d}\eta\right)}
+\frac{c_{2}(k)}{\sqrt{2ma^3}} \exp{\left(im\int a\, \text{d}\eta\right)}\,.
\label{wkb0}
\end{eqnarray}
For the normalized positive frequency solutions we get $c_1=1/(2\pi)^{3/2}$, $c_2=0$. Thus, for the energy density and pressure we obtain from \eqref{energydensity} and \eqref{pressure}
\begin{eqnarray}
\rho_{\text{ren}} = \frac{1}{8\pi^2} \int_0^{\Lambda_R} \text{d}k \, k^2\left( \frac{2m}{a^3}+\frac{9H^2}{4ma^3} + \frac{k^2}{ma^5} \right)\,
\label{energydensityarbitrary}
\end{eqnarray}
\begin{eqnarray}
p_{\text{ren}} =  \frac{1}{8\pi^2} \int_0^{\Lambda_R} \text{d}k \, k^2\left( \frac{9 H^2}{4m a^3} - \frac{k^2}{3ma^5} \right) \,.
\label{pressurearbitrary}
\end{eqnarray}
Thus, in this limit we get
\begin{eqnarray}
w=\frac{p_{\text{ren}}}{\rho_{\text{ren}}}\approx 0\,,
\end{eqnarray}
i.e. vacuum energy scales as non-relativistic matter  for
sufficiently heavy fields.
In terms of the comoving cutoff, the energy density reads
\begin{eqnarray}
\rho_{\text{ren}} = \frac{1}{12\pi^2}\frac{m}{a^3}\Lambda^3_{R}
\label{rren}
\end{eqnarray}
where we have only kept the dominant term. As it happens in other areas of physics, the cutoff value can be constrained observationally.
Thus, the amount of energy corresponding to the vacuum of this field cannot exceed the total amount of
energy of the Universe in its different stages. In particular, comparing (\ref{rren}) with the dark matter density, we obtain the following limit
on the comoving cutoff
\begin{eqnarray}
&&\Lambda_{R}\lesssim 8.96 \times 10^{-4}\, T_0 \left(\frac{m}{125\, \text{GeV}}\right)^{-1/3}\,,
\end{eqnarray}
where $T_0$ is the present temperature of the CMB and for the sake of concreteness we have normalized the scalar mass to the bosonic resonance recently measured at the LHC and compatible with the Standard Model Higgs field. Note that if the cutoff saturates the above inequality, the observed dark matter content of the Universe could be explained
as the vacuum energy corresponding to a particular scalar field. The  required comoving cutoff $\Lambda_R$ is a few orders of magnitude
smaller than the radiation temperature today and accordingly
the IR  condition ($ma\gg \Lambda_R$) is satisfied throughout the whole matter era for all the massive bosonic
fields in the Standard Model.

Let us then consider the effect of scalar perturbations on the
vacuum energy. We will
work in the longitudinal gauge, for which the perturbed metric reads
\begin{eqnarray}
\text{d}s^2 = a^2(\eta) \left\{ \left[1 + 2 \Phi(\eta,{\bf x})\right]\, \text{d}\eta^2 - \left[1 - 2\Psi(\eta,{\bf x})\right]\,\text{d}{\bf x}^2 \right\}\,.
\end{eqnarray}
The scalar field can also be expanded around the unperturbed solution as
\begin{eqnarray}
\phi(\eta,{\bf x}) =\phi_0(\eta,{\bf x})+\delta\phi(\eta,{\bf x})\,,
\end{eqnarray}
where $\phi_0$ satisfies
\begin{eqnarray}
\phi_0''+2\phi_0'{\cal H}-\nabla^2\phi_0+m^2a^2\phi_0=0\,,
\end{eqnarray}
and the total field can be shown to satisfy
\begin{eqnarray}
\phi''&+&(2{\cal H}-\Phi'-3\Psi')\phi'- (1+2(\Phi+\Psi))\nabla^2\phi\nonumber \\
&-&\boldsymbol{\nabla}\phi\cdot \boldsymbol{\nabla}(\Phi-\Psi)
+m^2a^2(1+2\Phi)\phi=0\,, \label{pert}
\end{eqnarray}
up to first order in perturbations, where $\cal H$ stands for $a'/a$.

In order to quantize the perturbed field, we will look for a complete orthonormal set
of  solutions of the above equation. For that purpose, we will try a WKB ansatz
for the solutions in the form
\begin{eqnarray}
\phi_k(\eta,{\bf x})=f_k(\eta,{\bf x})\, e^{i\theta_k(\eta,{\bf x})}\,
\end{eqnarray}
where $f_k(\eta,{\bf x})$ is a slowly evolving function of $\eta$ and ${\bf x}$ whereas
$\theta_k(\eta,{\bf x})$ is a rapidly evolving  phase.

The positive frequency solutions of the unperturbed equation with momentum ${\bf k}$
can be written as
\begin{eqnarray}
\phi^{(0)}_k(\eta,{\bf x})=F_k(\eta)\, e^{(i{\bf k}\cdot{\bf x}-i\int^\eta\omega(\eta')\,\text{d}\eta')}\,,
\end{eqnarray}
with $\omega^2=k^2+m^2a^2$ 
and
\begin{eqnarray}
F_k(\eta)=\frac{1}{a\sqrt{2(2\pi)^3\omega}}\,.
\label{Fsol}
\end{eqnarray}
in agreement with  (\ref{wkb0}), 
so that  we can expand the perturbed fields as
\begin{eqnarray}
f_k(\eta,{\bf x})&=&F_k(\eta)+\delta f_k(\eta,{\bf x})\,\nonumber \\
\theta_k(\eta,{\bf x})&=&-\int^\eta\omega(\eta')\,\text{d}\eta'+{\bf k}\cdot {\bf x}+\delta\theta_k(\eta,{\bf x})\,.
\end{eqnarray}
Substituting in (\ref{pert}) and expanding in metric perturbations, 
we get to first order
\begin{eqnarray}
\delta\theta'_k\simeq-ma\,\Phi\simeq -\omega\,\Phi
\end{eqnarray}
and
\begin{eqnarray}
\delta f_k=\frac{3F_k}{2}\Psi+\frac{F_k}{2}\nabla^2\int\left(\frac{1}{\omega}\int \omega\Phi\, \text{d}\eta\right) \text{d}\eta\,.
\end{eqnarray}
where again we have used  $m^2a^2\gg k^2$.

On the other hand, from the perturbed energy-momentum tensor we get
\begin{eqnarray}
\rho=\langle 0\vert T^{0}_{\;0}\vert 0\rangle&=& \int \frac{\text{d}^3{\bf k}}{2a^2} \left((1-2\Phi)\vert{\phi'_{k}}\vert^2+(1+2\Psi)\vert\boldsymbol{\nabla}\phi_{k}\vert^2 + m^2a^2\vert\phi_{k}|^2\right)\,
\end{eqnarray}
\begin{eqnarray}
p=-\langle 0\vert T^{i}_{\;i}\vert 0\rangle&=& \int \frac{\text{d}^3{\bf k}}{2a^2} \left((1-2\Phi)\vert{\phi'_{k}}\vert^2
-\frac{ (1+2\Psi)}{3}\vert\boldsymbol{\nabla}\phi_{k}\vert^2
- m^2a^2\vert\phi_{k}|^2\right)\,,
\end{eqnarray}
then for the density perturbation we get to first order
\begin{eqnarray}
\delta\rho&=& \int\text{d}^3{\bf k}\,2m^2F_k\,\delta f_k\nonumber \\
&=&
 \int_0^{\Lambda_R} \frac{\text{d}{k}}{2\pi^2a^3} k^2m\left(\frac{3}{2}\Psi+\frac{1}{2}\nabla^2\int\left(\frac{1}{a}\int a\Phi\, \text{d}\eta\right) \text{d}\eta\right) .
\label{densityp}
\end{eqnarray}
 We also see that the contributions to the pressure from the
kinetic  and potential  terms cancel each other so that
\begin{eqnarray}
\delta p&=& 0\,,
\end{eqnarray}
and the corresponding speed of sound is
\begin{eqnarray}
c_s^2=\frac{\delta p}{\delta \rho}= 0\,.
\end{eqnarray}

The $T^i_{\; j}$ components with $i\neq j$ can be neglected when compared to $T^0_{\;0}$
in the limit $ma\gg k$ which implies the  vanishing
of the anisotropic stress, i.e. $\Phi=\Psi$.

From (\ref{densityp}) and the background density (\ref{rren}), the density contrast reads
\begin{eqnarray}
\frac{\delta\rho}{\rho^{(0)}}=3\Psi+\nabla^2\int\left(\frac{1}{a}\int a\Phi\, \text{d}\eta\right) \text{d}\eta\,.
\end{eqnarray}
Notice that this result is independent of the cutoff and it  agrees with the standard expression for a hydrodynamical
fluid \cite{Mukhanov}. Thus in particular, in the matter dominated era,
a  fluid with $c_s^2=0$, has
$\Phi=\Psi=\text{constant}$, both on sub-Hubble and super-Hubble scales.
For sub-Hubble modes $\nabla^2\Phi\,\eta^2\gg \Phi$, thus 
the second term dominates and accordingly
\begin{eqnarray}
\frac{\delta \rho}{\rho^{(0)}}\propto a\,.
\end{eqnarray}
Thus, quite unexpectedly, on sub-Hubble scales, vacuum density perturbations could grow
very much in the same way as in cold dark matter models.

The  structure of vacuum studied in this work opens up fascinating possibilities 
 in the evolution of those perturbations which eventually could reach the non-linear regime and form bounded objects. Those objects, in the absence
of any supporting pressure,  would finally collapse in what we could call a 
vacuum-energy black hole. The study of these vacuum objects will deserve future consideration.

\vspace{0.5cm}
\subsection*{Acknowledgments}

This work has been supported by MICINN (Spain) project numbers FIS2011-23000, FPA2011-27853-01 and Consolider-Ingenio MULTIDARK CSD2009-00064.
FDA acknowledges financial support from the UAM+CSIC Campus
of International Excellence (Spain).




\end{document}